\documentstyle[epsfig]{elsart}

\begin{document}

\begin{frontmatter}

\title{Anatomy of three-body decay II. Decay mechanism and 
       resonance structure}

\author{E. Garrido} 
\address{ Instituto de Estructura de la Materia, CSIC, 
Serrano 123, E-28006 Madrid, Spain }

\author{D.V. Fedorov, A.S.~Jensen \and H.O.U. Fynbo}
\address{ Department of Physics and Astronomy,
        University of Aarhus, DK-8000 Aarhus C, Denmark }

\date{\today}

\maketitle

\begin{abstract}
We use the hyperspherical adiabatic expansion method to discuss the
the two mechanisms of sequential and direct three-body decay.  Both
short-range and Coulomb interactions are included. Resonances are
assumed initially populated by a process independent of the subsequent
decay.  The lowest adiabatic potentials describe the resonances rather
accurately at distances smaller than the outer turning point of the
confining barrier.  We illustrate with realistic examples of nuclei
from neutron ($^{6}$He) and proton ($^{17}$Ne) driplines as well as
excited states of beta-stable nuclei ($^{12}$C).
\end{abstract}

\end{frontmatter}

\par\leavevmode\hbox {\it PACS:\ } 21.45.+v, 31.15.Ja, 25.70.Ef

\maketitle

\section{Introduction.}

Resonance states consisting of a number of particles may decay into
final states of many fragments. Prominent examples are $\alpha$-decay,
nucleon emission and binary fission, where only two clusters are
present after the decay. The next step is three particles in the final
state which has been studied in various connections for many years,
see e.g. \cite{goo65}.  The recent years has witnessed enormous
progress in the treatment of the many possible three-body structures
\cite{jen04}.  As usual the continuum problem has turned out to be
more difficult, e.g. the first computation with the correct boundary
condition for decay into three charged particles is less than 10 years
old \cite{fed96}.  

The experimental techniques have developed tremendously the later years
and more details, accuracy and systematics are available for different
decays.  In particular the complicated structures, where both Coulomb
and short-range interactions are crucial, now begin to attract
increasing attention especially in the discussion of two-proton
radioactivity.  Experimental information accumulates both for ground
state two-proton decay of unstable systems along the proton dripline
\cite{kry95,chr02,bla03} and for excited states of more stable nuclei
\cite{bai96,fyn00,cam01,fyn03}. The analyses and interpretations of
the three-body decay experiments are essentially all consistent with
sequential decay \cite{bai96,fyn00,cam01,fyn03}. Direct decay into
three-body continuum states is in principle always possible and
perhaps preferred, for example for short-lived intermediate resonances
or when sequential decay is forbidden by energy conservation.

Different types of calculations with the focus on three-body decay
widths are now available, e.g. using $R$-matrix theory where only two
sequential two-body emissions is included \cite{bro03}, $R$-matrix
theory combined with microscopic shell model computations with the
inherent model restrictions \cite{bro02}, three-body models with
outgoing flux where essentially only direct decay to the continuum is
included \cite{gri03}, three-body models with Faddeev type of
components combined with complex coordinate scaling \cite{kur04}, or
from the Faddeev equations combined with either outgoing flux or
complex scaling \cite{gar04}.  The three-body models have very
restricted model spaces but are precisely tuned to describe three-body
structures.

Let us now consider a decaying resonance with complex energy where the
real and imaginary parts define the position and the partial
three-body decay width, respectively.  We assume that the decay is
independent of the formation as for lifetimes long compared to the
population process.  If the initial state is a many-body resonance the
three fragments in the final state must be created as part of the
decay process.  Using three-body models this opens for a definition of
three-body spectroscopic factors in analogy to the preformation
factors used for $\alpha$-decay \cite{fed02}. This tacitly assumes, as for
$\alpha$-decay, that the small distance many-body dynamics is
unimportant for the process. Similarly defined spectroscopic factors
can be computed with the shell-model \cite{bro03}, where a
non-stationary final state seems to be present after the decay..

The decay process leads from an initial to a final state of separated
particles. Outside the range of the strong interaction only the
Coulomb and centrifugal barriers remain. However, different paths to
the final-state of three free particles are still possible,
e.g. sequential decay or direct decay into the three-particle
continuum. In the preceding companion paper we discussed the relative
importance of these decay mechanisms in a schematic model where the
short-range interaction only is used to provide the correct resonance
energy \cite{gar05}.

The most advanced three-body models include both aspects in resonance
computations, where the correct asymptotic large distance boundary
conditions are properly accounted for by a complex energy or by
complex rotation \cite{fed96,myo01,gar02,fed03}.  The
intermediate configurations attempting to describe the process are
non-observables.  Virtual population of such states can only be
indicated through a model interpretation.  A distinction of decay
mechanism could then be rather uninteresting and perhaps even
impossible.  However, characterization of the reaction mechanism is
essential to understand a given process and indispensable for
generalizations to other processes, systems and observables.  In any
case a rigorous distinction between sequential and direct decays is
not meaningful without clear definitions.

The purpose of the present paper is to investigate the mechanism for
three-body decay from (possibly) many-body resonances.  We shall focus
on how the resonance wave functions are related to the decay
mechanisms which in turn produce for example the observable energy
spectra.  We shall report on elaborate computations for realistic
nuclear systems with both short and long-range interactions.  We
intent to bridge the gap between theory and the dominating
experimental analysis in terms of sequential decay.

\section{Resonances and the decay mechanisms}

Characteristic properties of a given system are revealed by the
complex poles of its $S$-matrix.  Poles of purely real, negative
energy on the physical and unphysical Riemann sheet correspond to
bound states and virtual states, respectively.  Poles with complex
energies, $E=E_r-i \Gamma/2$, correspond to resonances of energy $E_r$
and width $\Gamma$.  A peak structure of given position and width for
a specific observable can then be related to $E_r$ and $\Gamma$.  We
shall first compute energies for the poles of the resonances.  The
decay widths $\Gamma$ are then available and the corresponding wave
functions carry information about the decay mechanism. 

\subsection{General method}

We use the hyperspherical adiabatic expansion of the Faddeev equations
\cite{nie01} for the three particles appearing in the final state
after the decay. This means that we first must determine the interactions
$V_{ij}$ reproducing the low-energy two-body scattering properties.
Here the two-body bound and virtual states and the resonances are
possible fixpoints if the phase shifts are unavailable in the desired
region of energies.

A three-body short-range interaction depending on the coordinates of
all three particles may in addition be needed for example when the
precise resonance energy is required. This interaction may be very
small and amount to fine-tuning of the energy, but it could also be
substantial if the resonance at small distances is far from the
three-body structure appearing after the decay.  A relatively large
three-body interaction is usually needed when a many-body resonance
decays into three fragments.  

The effects of the three-body potential are changes of the (i) real
part of the energy to the desired value, which most effectively is
achieved with the strength, (ii) imaginary part of the resonance
energy achieved by reduction of the barrier, which is sensitive to the
range of the interaction, (iii) expansion coefficients of the radial
wave function on the adiabatic components, which only can be strongly
modified by a dependence on the quantum numbers of each adiabatic wave
function.  With a three-body potential which only depends on the
hyperradius, the short-range property then ensures that the decay
mechanism remains almost independent of strength and range. 

\begin{figure}
\begin{center}
\vspace*{-1.1cm}
\epsfig{file=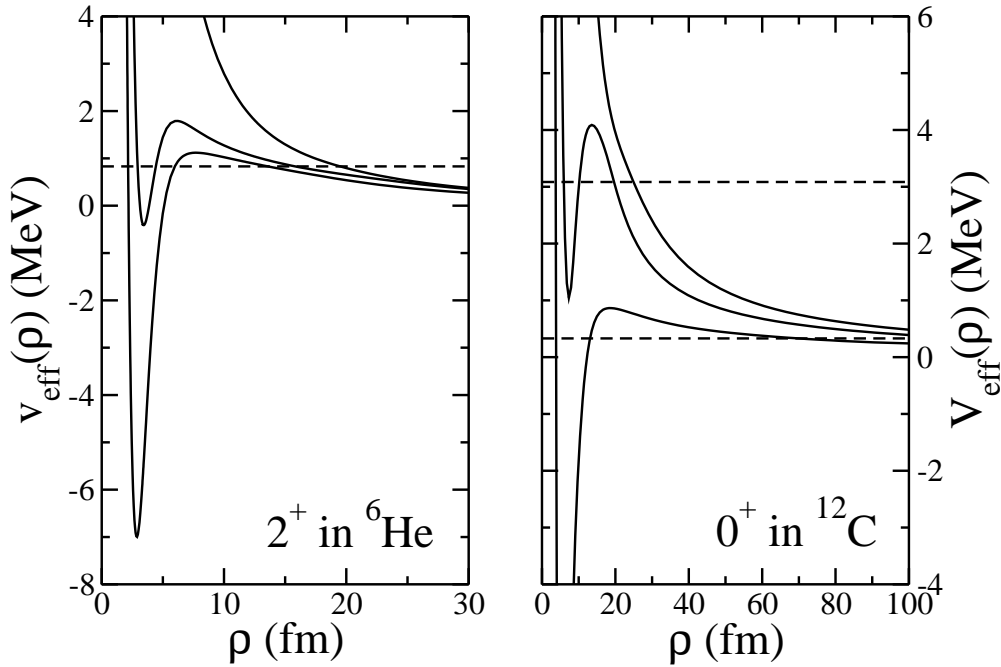,scale=0.5, angle=270}
\end{center}
\caption{ The lowest adiabatic potentials as functions of $\rho$. Left 
side shows the 2$^+$ states in $^{6}$He ($^{4}$He + n + n). The
resonance energy (horizontal line) and the width are 0.824(25) MeV and
0.113(20) keV \cite{til02}.  Right shows the 0$^+$ states in $^{12}$C
($\alpha + \alpha + \alpha$).  For the lowest two unbound states the
energies (horizontal lines) and widths are, 0.369(10) MeV and 8.5(1.0)
eV, 3.0(3) MeV and 3.0(7) MeV, see \cite{ajz90}.}
\label{fig1}
\end{figure}

We use hyperspherical coordinates as for example defined in
\cite{gar05}.  For fixed hyperradius, $\rho$, we then first solve the 
angular Faddeev equations to obtain the adiabatic potentials, see
examples for the diagonal parts in Fig.~\ref{fig1}.  All these
potentials diverge for small $\rho$ due to the generalized centrifugal
barrier and vanish for large $\rho$ at least as $\rho^{-3}$ for the
short-range interaction of $^{6}$He and as $1/\rho$ for the Coulomb
potential for $^{12}$C \cite{nie01}.  At relatively small distance the
attraction leads to minima responsible for bound states and resonance
structures.

The total wave function is expanded on the angular eigenfunctions
where the $\rho$-dependent expansion coefficients are the hyperradial
wave functions.  They are obtained from the resulting set of coupled
hyperradial equations solved with the boundary conditions appropriate
for the problem under investigation.  Virtual states and resonances
can then in principle be found by the conditions of outgoing fluxes in
all channels.  This requires a complex energy which then is the
resonance position and width.  The precise computation with this
complex energy method is in practice numerically delicate
\cite{cob97}.

Another much more efficient method is to ``rotate'' all the
coordinates by the same angle into complex values like $\rho
\rightarrow \rho \exp(i\theta)$ \cite{fed03}. Since all other coordinates 
are angles, i.e. ratios of lengths, only $\rho$ becomes complex. If
$\theta$ is sufficiently large both bound states and resonances are
obtained with the bound state boundary condition of exponential
fall-off of the radial wave function for large hyperradius.  The
resulting resonance energies are again complex numbers corresponding
to positions and widths.  For bound states any rotation $\theta$ can
be used.  One adiabatic potential is sometimes sufficient to describe
the resonance without any complex rotation.

\subsection{Resonance decay mechanisms} 

The full computation of $S$-matrix poles provides widths of the
resonances.  However, this quantum mechanical partial three-body decay
width only gives the weighted sum of all amplitudes leading
from the initial to the final state.  The dominating path(s) defining
the decay mechanism(s) are not revealed in the value of the width.
Fortunately the model does contain such information \cite{fed96} and
we shall try to extract it.

The picture of sequential decay is very often used in data analyses
\cite{kry95,chr02,bla03,fyn03}.  The measurements provide the relative 
energy between pairs of fragments, i.e. the excitation energy of the
corresponding two-body subsystem. The experimental definition of
sequential decay is that this energy, properly extrapolated to small
distances, matches a known resonance energy.  The outcome is
frequently that sequential decay is overwhelmingly dominating.  This
is perhaps not surprising since no other mechanism is included.
However, careful comparison between experiment and $R$-matrix theory
suggest that contributions of a different origin sometimes seem to be
necessary \cite{bar03}.

At this point it may be appropriate to emphasize that any complete
basis can be used to describe these decay processes occurring for
positive energies in the continuum.  A complete basis could be product
functions of two complete two-body basis sets of the relative motion
of (i) two particles and (ii) their center of mass and the third
particle.  This choice matches perfectly with the sequential decay
process.  However, other complete basis sets are possible for example
the set of continuum three-body wave functions which includes the
three-body resonances, or the rotated wave functions where the
resonances explicitly are separated out \cite{myo01}.  The
latter basis sets are more appropriate when the decay directly
populates states which are complicated superpositions of the simple
sequential decay basis.  The best choice is not a matter of principles
or decay mechanism, but a practical question of faster or slower
convergence.

The adiabatic expansion method provides a complete basis consisting of
the angular eigenfunctions.  Each of these adiabatic eigenfunctions is a
function of the hyperradius varying from small to large values,
i.e. leading from initial to final state by a fully specified
amplitude provided by the self-consistent adiabatic adjustment of the
particles as the average size (hyperradius $\rho$) increases. This path
could for example describe the two steps of a sequential decay
process.  However, it could also be a specific coherent superposition
of more than one sequential process. It could also be something else
like all particles proportionally increasing their mutual distances,
or any other coherent or incoherent combination of relative motion.
Clearly more combinations become possible when more than one adiabatic
component contribute, but the principle is still the same, i.e. a
specific weighted combination of paths from initial to final state.
The great advantage is that the adiabatic eigenfunctions by definition
are optimum choices for each distance.

Sequential decay seems intuitively most likely when a narrow two-body
resonance offers an intermediate stepping stone.  In turn this is most
appropriate when the initial three-body resonance wave function has a
very large overlap with the two-body resonance wave function
multiplied by some function depending on the relative coordinate of
the last particle.  Then the decay towards larger distances would
proceed in this configuration until other channels are populated. This
typically occurs when other adiabatic potentials corresponding to
different configurations cross.  The crucial sizes of the related
couplings are closely connected to the imaginary part of the two-body
$S$-matrix pole which is half the width. The qualitative relation is
that a small width means small coupling and therefore pronounced
sequential decay through the initial two-body resonance.

In contrast direct decay is expected to dominate both for a large
two-body resonance width implying strong couplings to other channels
and when the initial three-body resonance wave function has no
configuration similar to a two-body subsystem in a resonance. In the
first case the decay path would quickly change away from the
sequential behavior and not come back. In the latter case the decay
path can only change character and look like sequential decay if the
adiabatic potentials couple strongly to precisely such a
configuration. This is much more unlikely since many other channels
also are open directly to the continuum.

The hyperspherical adiabatic expansion method includes simultaneously
all decay channels, i.e. provides automatically the partial three-body
decay width when the decay is either direct or sequential or any
combination.  Unlike other approaches no assumption is a priori made
about the decay mechanism.  All channels can be obtained in one
computation.  Thus we have the prerequisites for discussing the
mechanism of three-body decay of a given resonance. The task could be
to characterize different decay mechanisms and investigate under which
conditions the corresponding paths are dominating.  In particular, we
would like to know if sequential decay dominates over direct decay or
vice versa.

\section{Realistic calculations}

The schematic model with only Coulomb potential and centrifugal
barrier provides indication of the dominating decay mode \cite{gar05}.
However, the short-range interaction is often decisive, i.e.  when
correlated intermediate configurations along the decay path minimize
the dominating action integral.  The adiabatic potentials are
precisely constructed to carry such signature of the three-body
structure as the system expands.  Each potential could correspond to
one dominating configuration, e.g. sequential decay where one two-body
subsystem is in a favored resonance state while the third particle
moves away. Different types of sequential decay can exist, i.e. via
different two-body resonances in the same subsystem (coherent) or via
different subsystems (incoherent).  With one dominating potential the
width can be reliably estimated by the (WKB) tunneling width.  We
shall illustrate by realistic examples.

\subsection{Short-range potentials: $^{6}$He(2$^+$)}

\begin{figure}
\begin{center}
\vspace*{-1.1cm}
\epsfig{file=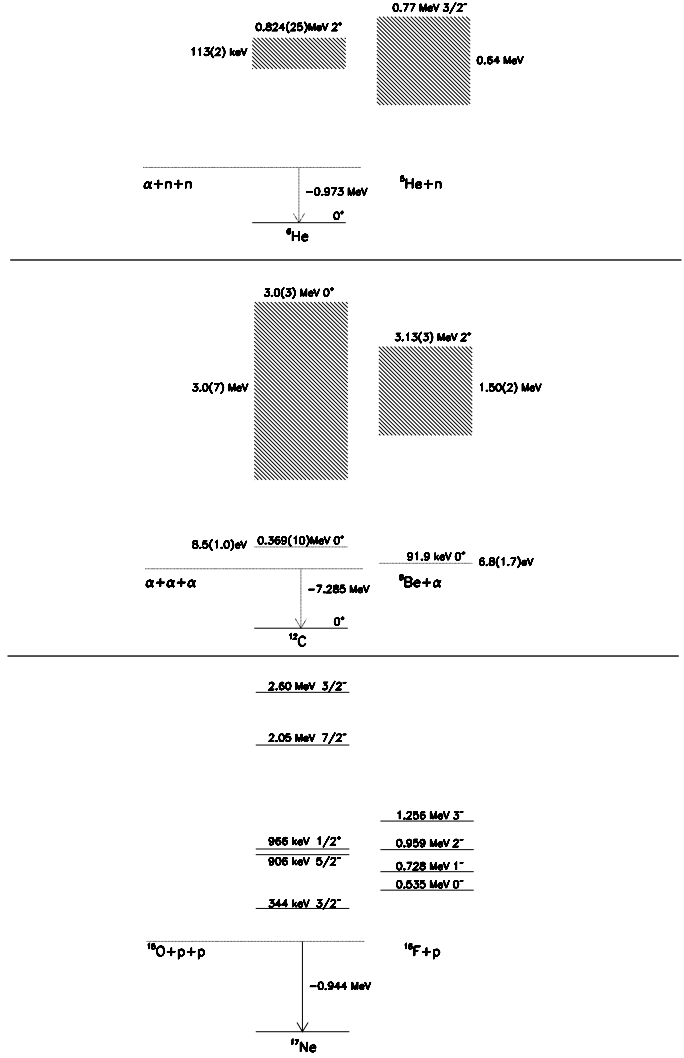,height=14cm}
\end{center}
\vspace*{0.0cm}
\caption{The experimental low-energy spectra for $^{5}$He, $^{6}$He,
$^{12}$C, $^{8}$Be, $^{17}$Ne and $^{16}$F. The data are from \cite{til02},
\cite{ajz88}, \cite{gui98}. }
\label{fig10}
\end{figure}

The first example is the well studied two-neutron Borromean halo
nucleus, $^{6}$He ($^{4}$He+n+n), where details of interactions and
computational accuracy are available. Furthermore the analytically
established asymptotic behavior of the adiabatic potentials is reached
at relatively small distances since only the short-range interaction
is present.  Both the 0$^+$ ground state and the 2$^+$ excited state
are rather well described as three-body states, see Fig.~\ref{fig10}.
The two-body subsystem of n+$^{4}$He has no bound states but a
$p_{3/2}$-resonance, and the two-neutron system has a virtual state at
$-0.14$~MeV \cite{gar02}.

\begin{table}
\caption{For the $2^+$ resonance in $^6$He, the two lowest $0^+$ resonances in 
$^{12}$C, and the 3/2$^-$ and 5/2$^-$ resonances in $^{17}$Ne we give the 
corresponding energies and widths computed with the realistic calculations
described in \cite{gar04,fed03} (third and fourth columns), the width in 
the WKB approximation as described in \cite{gar04} (fifth column), and using
the schematic approaches assuming direct and sequential decay given in
\cite{gar05} (sixth and seventh columns). In the fourth row, the number in
brackets corresponds to sequential decay through the 2$^+$ state in $^8$Be.
The last two columns are the experimental values. All the energies and widths 
are given in MeV.}
\label{tab1}
\begin{center}
{\scriptsize 
\begin{tabular}{|c|cccccccc|}
\hline
  & $J^\pi$ &$E_{real.}$ & $\Gamma_{real.}$ &
$\Gamma_{WKB}$ & $\Gamma_{sch.}^{(Dir.)}$ &
$\Gamma_{sch.}^{(Seq.)}$ & $E_{exp.}$ & $\Gamma_{exp.}$  \\ \hline
$^6$He & $2^+$ & 0.82 & 0.12 & 0.19 &0.01 & 0.25 & 0.84 & 0.09 \\ \hline
$^{12}$C & $0^+_1$ & 0.33 & $2\cdot 10^{-5}$ & $6\cdot 10^{-5}$ 
                & $\sim 10^{-6}$ & $\sim 10^{-4}$ & 0.37 & $8\cdot 10^{-5}$ \\
& $0^+_2$ & 4.3 & 0.64 & 0.4 & 0.2 & 0.7 (0.3) & $3.0\pm0.3$ & $3\pm0.7$ \\ \hline
$^{17}$Ne & $\frac{3}{2}^-$ & 0.34 & --- & $3.6\cdot 10^{-12}$ & $\sim 10^{-9}$ &
--- &  0.34 & $<2.5\cdot 10^{-11}$ \\
 & $\frac{5}{2}^-$ & 0.82 & --- & $1.3\cdot 10^{-10}$ & $\sim 10^{-5}$ &
$\sim 10^{-5}$ &  0.82 & $>3\cdot 10^{-10}$ \\ \hline 
\end{tabular}
                           }
\end{center}
\end{table}

Then (parts of) the decay of the 2$^+$-state could be sequential via
the broad $^{5}$He-resonance or via $\alpha$-emission leaving a
two-neutron configuration behind.  The lowest adiabatic potentials are
shown in Fig.~\ref{fig1}. The different numerical results are summarized
in the second row of table~\ref{tab1}. The full computation from the complex
rotation method with (without) the weak three-body potential gives a
resonance energy of 0.82 MeV (1.1~MeV) and a width of 0.12 MeV, which
is remarkably close to the measured values \cite{ajz88}. These values
are quoted as realistic in the third and fourth columns of table~\ref{tab1}.
Using only the lowest non-rotated adiabatic potential with the energy equal to
0.82 MeV the WKB estimate for the width \cite{gar04} is 0.19 MeV (fifth column),
where the uncertainty in the knocking rate alone can explain the deviation 
from 0.12~MeV. Therefore the wave function along the path defined by this
potential reveals the decay mechanism by showing the structure
continuously changing from small to large distances. In table~\ref{tab1}
we also show the results corresponding to the schematic approaches for a 
centrifugal barrier potential as given in \cite{gar05} (sixth and seventh
columns). They refer to the direct decay and the sequential decay through
the $p_{3/2}$ resonance in $^5$He. In these calculations we have used 
a ratio between the outer and inner turning points of around 3, as suggested
by the lowest adiabatic potential for this resonance shown in Fig.~\ref{fig1}.
The knocking rate multiplying the transmission coefficient is put equal to a 
typical value of 3 MeV$/\hbar$. These estimates suggest that the decay 
of the 2$^+$ resonance in $^6$He could be sequential. However, this result
is highly sensitive to the value of the ratio between the turning points, 
that for sequential decay could be very different from direct decay.
A ratio equal to 5 gives an estimated width for the sequential decay similar 
to the one quoted for the direct decay.

\begin{figure}
\begin{center}
\vspace*{0.1cm}
\epsfig{file=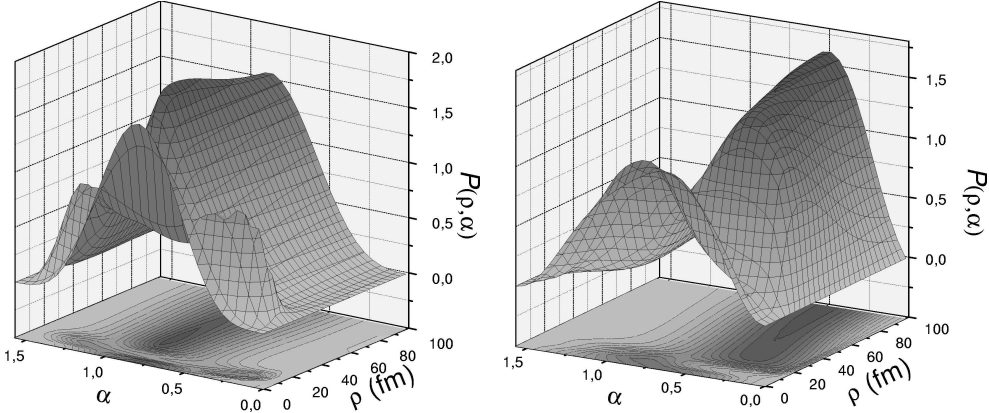,scale=0.6,angle=270}
\end{center}
\caption{ The probability distribution for $^{6}$He(2$^+$) for the lowest 
adiabatic potential as function of hyperradius $\rho$ and $\alpha$
related to the distance by $r_{ik} \propto \rho \sin \alpha$, i.e. the
distance between either the one neutron and core $r_{nc}$ (left) or
the two neutrons $r_{nn}$ (right).  }
\label{fig3}
\end{figure}

The probability distributions shown in Fig.~\ref{fig3} are obtained by
integration over all coordinates except the hyperradius and the
distance between two of the particles. When the distance refers to one
neutron and the $\alpha$-particle we see three peaks for small $\rho$,
i.e.  at small and large $\alpha$ corresponding to an
$\alpha$-particle spatially close to one of the neutrons and an
intermediate $\alpha \approx \pi/4$ corresponding to a triangular
configuration with equal distances between all three particles.

As $\rho$ increases through and far beyond the barrier the probability
appears to peak for the triangular configuration.  Sequential decay
through the two-body resonance can not be seen. This is consistent
with the rather low barrier height observed in Fig.~\ref{fig1}, which
does not leave any room for a flat region above the two-body resonance
at 0.77 MeV. The generalized centrifugal barrier is already lower when
$\rho$ is around 15~fm.  Thus the short-range interaction and the
broad two-body resonance seems to make it advantageous to take the
direct decay road to the continuum.  The two-body energy is also
rather close to the three-body resonance energy, but the main reason
is that the third particle can feel the short-range interaction until
distances outside the low and thin barrier.

To illustrate the other possible sequential decay mode of
$\alpha$-emission we also show in Fig.~\ref{fig3} the probability as
function of $\rho$ and the distance between the two neutrons.  At
small $\rho$ the dominating feature is a peak corresponding to roughly
equal distance between the two neutrons and their center of mass and
the $\alpha$-particle.  This distribution reflects the same initial
triangular resonance structure as seen in the other coordinate system
in Fig.~\ref{fig3}.  As $\rho$ increases the probability peaks at
relatively small values of the $\alpha$-coordinate.  However, applying
the proper mass scaling we find that the distance between the two
neutrons is comparable to the neutron-$\alpha$ distances.  Thus from
both plots we find that all three particles on average move apart with
roughly equal distance between all pairs.  This is compatible with
direct decay into the three-body continuum. Neither the $s$-wave
attraction (virtual state) between the neutrons nor the $p$-wave
resonance in the $n-\alpha$ system is capable of producing a
substantial sequential component.

\begin{figure}
\begin{center}
\vspace*{-1.1cm}
\epsfig{file=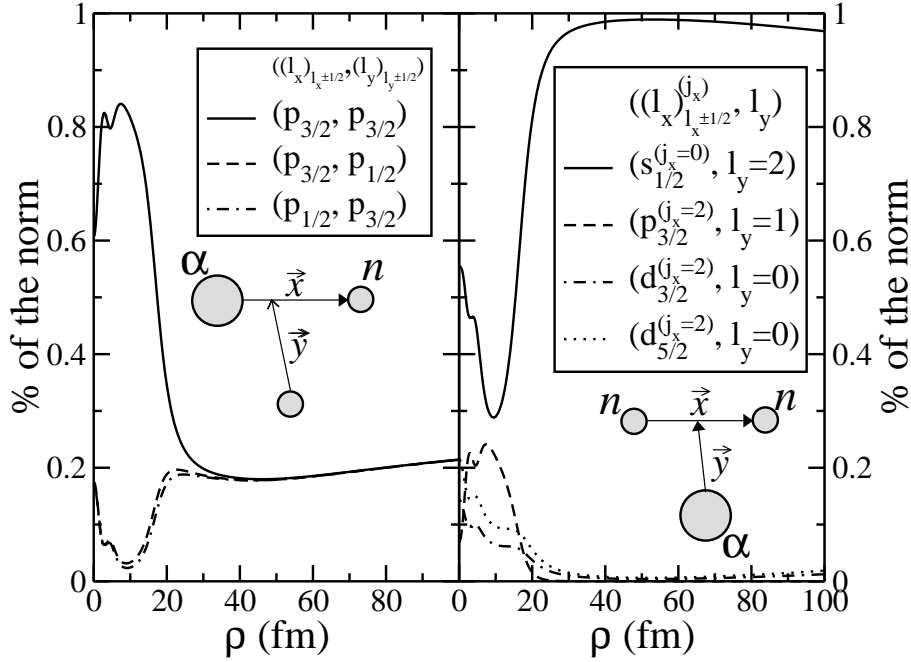,scale=0.5, angle=270}
\end{center}
\caption{The fraction of different components in the lowest adiabatic 
potential as function of $\rho$ for $^{6}$He(2$^+$). The angular
momenta are specified by $\ell_x$, $j_x$, $\ell_y$, $j_y$ and
$J=2$. Left $x$ refers to the neutron-$\alpha$ system and $y$ to its
center of mass motion relative to the other neutron. We give the
$(x,y)$ components on the figure as $\ell_{j}$. Right $x$ refers to
the two-neutron system and $y$ to its center of mass motion relative
to the $\alpha$-particle. We give the $(x,y)$ components on the figure
as $\ell_{j}^{(j_1)}$, where $j_1$ is the angular momentum quantum
number obtained by coupling of $\ell_x$ and the neutron spin of $1/2$.}
\label{fig4}
\end{figure}

The structure of the wave function is characterized by the quantum
numbers of its different components.  We show in Fig.~\ref{fig4} how
they change as $\rho$ increases. By far the largest component at small
distances is a $p_{3/2}$ neutron-$\alpha$ structure coupled to the
other neutron in precisely the same relative state around the common
center of mass.  This structure is maintained in the non-classical
region, but changes quickly outside the barrier where all three
possible different $p$-wave combinations become equally probable.
This again confirms the previous observation that no two-body
subsystem is used as a stepping stone on the path to full separation
of all particles. If for example the $p_{3/2}$ resonance first would
be populated the $p_{1/2}$ neutron-$\alpha$ component should have been
smaller at the large distances. Now instead all particles
simultaneously move outside the range of the short-range interaction
and preference for such fine-structure is lost.

It may be informative to express the components in terms of a
different coupling scheme where the relative motion of the two-neutron
system first is established and then coupled to the motion of the
$\alpha$-particle. As seen in Fig.~\ref{fig4} at small distances a
number of components of comparable magnitudes are present, but
immediately outside the barrier and at larger distances only one
component remains, i.e. the two neutrons in relative $s$-states
coupled to angular momentum zero moving around the $\alpha$-particle
in a $d$-state.  This structure could of course be called a
dineutron-$\alpha$ $d$-state, and the decay could correspondingly be
mistaken for a sequential decay via emission of an
$\alpha$-particle. However, no intermediate two-body structure is
populated.  Instead all distances increase proportionally until the
decay is completed and all particles are free.  This is because the
short-range interaction already lowered the barrier for direct
emission to a lower energy than for the intermediate configuration.

\subsection{Symmetric cases with Coulomb potential: $^{12}$C(0$^+$)}

A relatively simple Coulomb dominated structure is found in the
celebrated second $0^+$-state in $^{12}$C which approximately can be
described as a three-body resonance \cite{fed96}.  The two-body
subsystem can then only be $^{8}$Be with the $0^+$ ground state, see
Fig.~\ref{fig10}.  These narrow two and three-body widths are both due
to prominent Coulomb barriers.  The higher-lying $0^+$-state is
computed to be at the energy 4.3~MeV with a width of 0.6~MeV
\cite{fed03}. It could then decay sequentially via the $2^+$ 
rotational-like state in $^{8}$Be at about 3 MeV.
 
To study the decay mechanism we again use the lowest adiabatic
potentials shown in Fig.~\ref{fig1}. The fine-tuning with the
three-body potential reproduces the measured resonance energy and the
width approximately in full computations with several adiabatic
potentials and complex rotation.  Using only the lowest non-rotated
adiabatic potential with the correct energy of 0.37 MeV for the lowest
$0^+$-state we find the WKB width of about 0.060~keV, i.e. three times
larger than the full computation \cite{fed96} and about eight times
larger than the measured value.  For the next $0^+$ resonance the
experimental energy of 3.0 MeV leads to a WKB width around 8~keV which
is much smaller than the experimental value of 3.0~MeV.  The computed
energy of 4.3 MeV leads instead to the much larger WKB width of 0.4
MeV, because now the two turning points are very close to each other
and the barrier penetration is much more probable.  This width is
comparable to 1.1~MeV found in \cite{kur04} where the inconsistency
between position and width also was noticed.  It is encouraging that
this larger computed energy of about 4~MeV is in better agreement with
recent experimental values \cite{fyn04}. These results are collected
in the third and fourth rows of table~\ref{tab1}. 

The results
corresponding to the schematic calculations described in \cite{gar05}
are also given. These estimates depend only on the ratio between the
outer and inner turning points. This ratio is taken equal to 4 and 2 for 
the first and second $0^+$ resonances, respectively. These values are
suggested from the effective potential for each of the two resonances 
shown in the right part of Fig.~\ref{fig1}. The knocking rate is again
taken equal to 3 MeV/$\hbar$. For the second $0^+$ resonance
two estimates for the case of sequential decay are given, one for
decay through the lowest 0$^+$ state in $^8$Be, and also, in parenthesis,
the one assuming sequential decay through de 2$^+$ resonance in $^8$Be. 
The absolute values from the schematic model deviate by less than about 
one order of magnitude from the experimental values.

\begin{figure}
\begin{center}
\vspace*{-1.1cm}
\epsfig{file=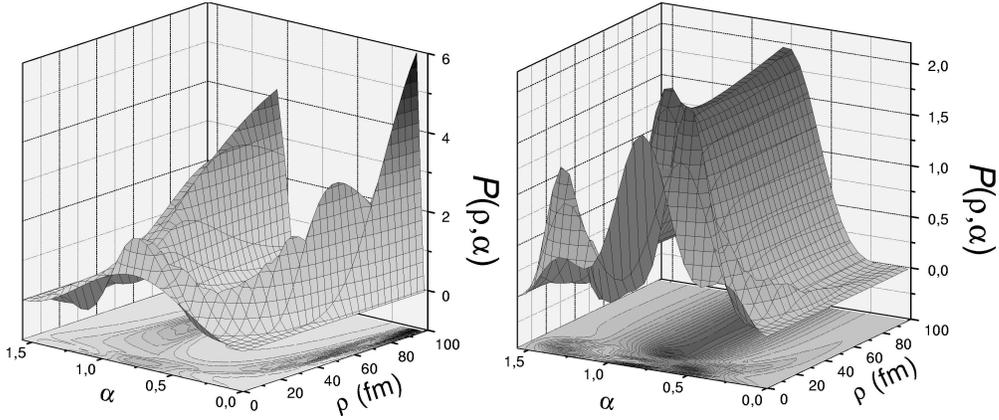,scale=0.6, angle=270}
\end{center}
\caption{ The probability distribution for $^{12}$C(0$^+$) for the first
(left) and second (right) adiabatic potential as function of
hyperradius $\rho$ and $\alpha$ related to the distance between two
$\alpha$-particles. }
\label{fig5}
\end{figure}

The lowest $0^+$ resonance wave function has a probability of more
than 95~\% arising from the first adiabatic potential.  We show in
Fig.~\ref{fig5} its structure revealed by the probability distribution
as function of hyperradius and distance between two
$\alpha$-particles.  At small $\rho$ the dominating peak is at $\alpha
\approx \pi/4$ consistent with a diffuse elongated structure. As $\rho$ 
increases this peak is divided into two separated ridges, i.e. one at
very small $\alpha$ corresponding to two close-lying particles and one
at $\alpha \approx 1$ consistent with the configuration arising from
symmetrization of this classical structure.

The probability distribution reflects sequential decay as expected
from a decay where the short-range attraction can be exploited until
one Coulomb interaction is strongly reduced by removal of one of the
charged particles.  This sequential decay mechanism is only favored
due to the short-range interaction which is responsible for the ground
state structure of $^{8}$Be.  Otherwise direct decay would have been
as probable as discussed for schematic models \cite{gar05}.  This
conclusion is strongly depending on the total energy of the decay and
its possible division into the two sequential steps.

The second excited $0^+$-resonance has a probability of more than
98~\% arising from the second non-rotated adiabatic potential. The
corresponding probability distribution is also shown in
Fig.~\ref{fig5}. At small $\rho$ we now see three peaks revealing a
different structure. Nevertheless, as $\rho$ increases only one
prominent ridge remains at $\alpha \approx \pi/4$ corresponding to
equal scaling of all distances between pairs of particles. This direct
decay mechanism is not overshadowed by sequential decay via the
excited $2^+$ rotational-like state in $^{8}$Be at about 3 MeV.
Apparently the short-range interaction cannot compete when the penalty
for angular momentum first must be payed.

\begin{figure}
\begin{center}
\vspace*{-1.1cm}
\epsfig{file=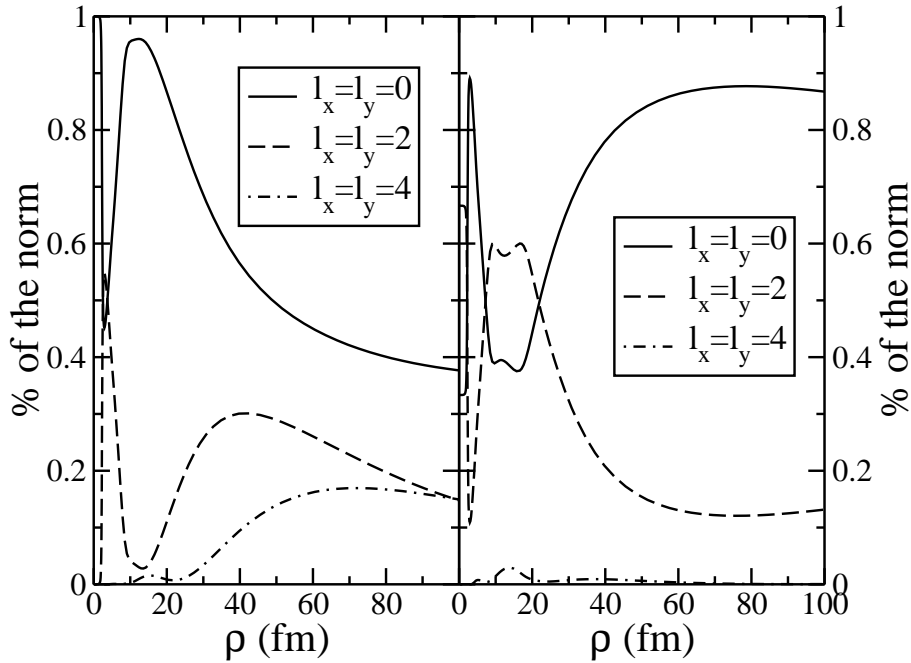,scale=0.5, angle=270}
\end{center}
\caption{The fraction of different components in the first (left) and
second (right) adiabatic potential as function of $\rho$ for
$^{12}$C(0$^+$). The partial orbital angular momenta are equal since
they couple to a total of $0$ and all intrinsic spins are zero.  }
\label{fig6}
\end{figure}

The structure of the resonance wave functions can be seen in
Fig.~\ref{fig6} where the quantum numbers of the components are given
as function of $\rho$.  The quickly changing contribution at small
$\rho$ is due to coupling and subsequent admixtures of the two
close-lying adiabatic potentials in this region. For the first
$0^+$-resonance the partial $s$-waves dominate for ground state
distances as also found in \cite{kur04}. As $\rho$ increases beyond
about 25~fm the $s$-wave contribution decreases substantially while
correspondingly the higher partial waves contribute more as the
sequential decay structure is changing into the structure of three
free $\alpha$-particles. This is apparently achieved most efficiently
by exchanging angular momenta between the subsystems.

In comparison the second $0^+$-resonance has a very different behavior
as function of $\rho$. The small $\rho$-admixture has the same fast
variation, but in the intermediate region $s$ and $d$-waves are about
equally probable. The $d$-waves of about 50\% in the ground state are
apparently smaller than in \cite{kur04} where they dominate.  For the
decay process this variation may be interpreted as two simultaneously
contributing sequential decays via $0^+$ and $2^+$-states in $^{8}$Be.
However, this interpretation is not comparable with the probability
distribution in Fig.~\ref{fig5}. At larger distances the exchange of
angular momenta is no longer needed in this direct decay and the
$s$-wave probability quickly increases towards unity.  All distances
scale proportionally and all partial waves can as well assume their
lowest possible values reflected in the decrease of $d$-waves while
all higher partial waves remain small.

Even if angular momentum $2$ is populated at intermediate distances
sequential decay through the $2^+$ state in $^{8}$Be is not
automatically preferred. The large two-body width indicates direct
decay since an $\alpha$-emission quickly would be followed by a
two-body decay occurring already at distances comparable to the
$^{12}$C radius \cite{gar05}.  The dominating $s$-wave component at
larger distances also contradicts $2^+$ sequential decay, but allows
the conclusion from experiments of dominating $0^+$ sequential decay
\cite{sch66}.  The measured energy spectra still seems to contradict
the direct decay as exhibited by the probability distribution in
Fig.~\ref{fig5}.  One complication is that the coherent population of
the two $0^+$ resonances in experiments produce inseparable
contributions. However, the fast increase of the $l=0$ component probably 
must be interpreted as sequential decay if the wave function is expanded 
on a basis of $^8$Be-states multiplied by distorted waves. This is implicit 
in the usual analysis using R-matrix theory.

\subsection{Asymmetric cases with Coulomb potential: 
$^{17}$Ne($^{15}$O+p+p) resonances}

Beyond or at the proton dripline peculiar structures are possible,
because the Coulomb repulsion can create a barrier which confines
otherwise unbound charged particles to remain at short distances in
states of long lifetimes.  The lightest Borromean two-proton nucleus
$^{17}$Ne has several such resonance states, see Fig.~\ref{fig10}.
The ground state and the two first excited states are all essentially
of three-body structure \cite{gar04}.  The Borromean nature prohibits
proton decay from the ground state but also the $3/2^-$ state has too
low an energy to allow proton emission. In contrast the $5/2^-$ state
has sufficient energy to emit a proton and a two-step sequential decay
in its pure form is in principle allowed, see Fig.~\ref{fig10}.

\begin{figure}
\begin{center}
\vspace*{-1.1cm}
\epsfig{file=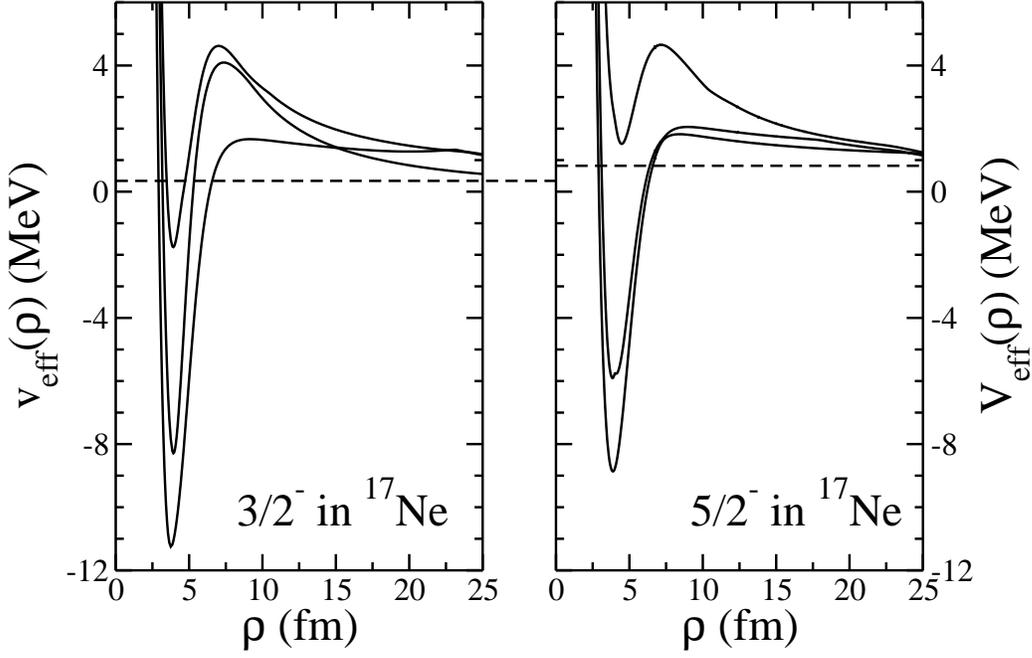,scale=0.5, angle=270}
\end{center}
\caption{ The lowest adiabatic potentials as function of $\rho$ for 
the states of $\frac{3}{2}^-$ (left) and $\frac{5}{2}^-$ (right) in
$^{17}$Ne($^{15}$O+p+p).  The resonance energies at 0.34~MeV and
0.82~MeV are indicated by the horizontal lines.}
\label{fig7}
\end{figure}

The lowest adiabatic potentials are shown in Fig.~\ref{fig7} as
functions of $\rho$ for $3/2^-$ and $5/2^-$.  The $3/2^-$ resonance is
dominated by the lowest-lying potential at small distances. This
potential has the remarkable behavior of being very flat in a
relatively large region, reaching at least to $\rho=25$ fm after
crossing other potentials.  A two-body substructure with the third
particle far away would produce this behavior. The slow decrease for
$\rho$ larger than about 8~fm has to be present due to the long-range
Coulomb repulsion between the third particle and the subsystem.  Also
the decreasing centrifugal barrier potential should be visible in this
potential.  The other adiabatic potentials decrease more regularly as
expected when the short-range interaction has vanished.  These
structures then correspond to configurations where all three particles
are spatially separated.

\begin{figure}
\begin{center}
\vspace*{-0.1cm}
\epsfig{file=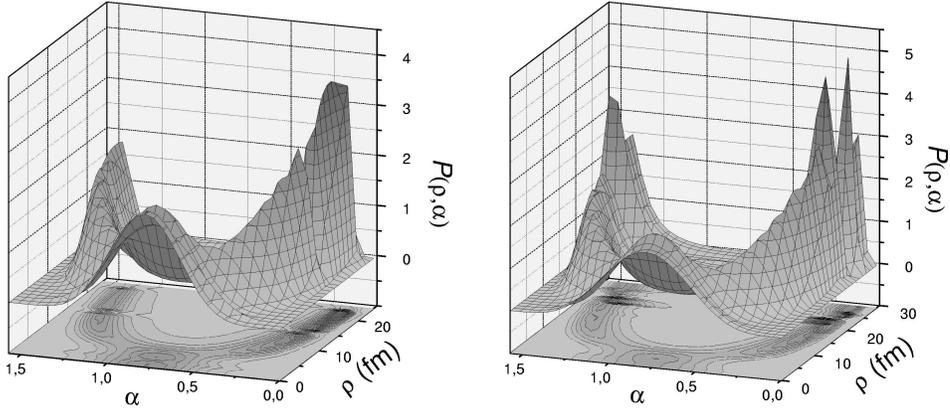,scale=0.6, angle=270}
\end{center}
\caption{ The probability distribution for the dominating adiabatic 
potentials of $^{17}$Ne($\frac{3}{2}^-$) (left) and
$^{17}$Ne($\frac{5}{2}^-$) (right) as function of hyperradius $\rho$
and $\alpha$ related to the distance between the proton and the
$^{15}$O-core.  }
\label{fig8}
\end{figure}

The two-proton decay of the $3/2^-$ state cannot be strictly
sequential, since the energy is below the energy of the unstable
ground state of $^{16}$F.  This does not prevent the decay from
proceeding via the energetically favorable path described by the
lowest adiabatic potential until at some point energy conservation
eventually dictates that also this two-body structure must be
broken. The decay mechanism is then best described as virtual
sequential decay through one of the $^{16}$F states.  This mechanism
is equivalent to the decay through the tail of an energetically
unaccessible state (a ghost) described by other authors to contribute
significantly \cite{bro02}.  The corresponding probability
distribution is shown in Fig.~\ref{fig8}. At small distance only one
prominent peak is present but quickly this develops into two ridges
at small and large distances between one proton and the core. The
ridge at large $\alpha$ corresponds to the large distance between core
and emitted proton.

We know that the resonance at small distance is dominated by
components where the proton-core two-body configurations are of
$sd$-structure \cite{gar04}.  Two $s$-waves cannot couple to $3/2$ due
to antisymmetry of the two protons.  The angular momentum of $3/2$ is
reached when both $s$ and $d$-resonances in $^{16}$F are maximally
exploited.  Two decay paths seem to be advantageous, i.e. sequential
decay via either the $s$-state at 0.53~MeV or the $d$-state at
0.95~MeV in $^{16}$F.  In both cases the first step would only amount
to removal of one proton while the other is left in the same state.
The energies of these intermediate configurations differ by 0.42~MeV,
but also by the centrifugal potential for the relative motion between
the emitted proton and the $^{16}$F $s$ or $d$ two-body state. Since
the $s$-state is lowest the centrifugal barrier from the additional
two units of $\hbar$ compensates at intermediate distances almost
precisely for the energy difference of 0.42~MeV.

The potential energy of the flat region is about 0.25~MeV larger than
obtained by adding Coulomb and centrifugal contributions to the
two-body states of the corresponding $^{16}$F configurations.  Thus
the intermediate state cannot be precisely the structures of the
$^{16}$F-resonances.  The most crucial decay path is via the
dominating potential followed through the crossing at 15~fm up to
$\rho= 25$~fm and then taking the lowest potential to the classical
turning point.  The corresponding WKB estimate (see table~\ref{tab1}) for the 
width obtained is found to be $\simeq 3.6 \times 10^{-12}$~MeV consistent 
with the experimental constraints from the $\gamma$-decay \cite{chr02,gar04}.
In contrast this is far from the estimate of $\simeq 1.2 \times
10^{-19}$~MeV obtained in \cite{gri03} where the method essentially
excludes contributions from the virtual sequential decays, which is
overwhelmingly dominating in our width computation.  Furthermore, the
chosen structure of the spin-dependent two-body interaction in
\cite{gri03} produces inseparable admixtures of proton-core $d_{3/2}$ and
$d_{5/2}$-states within the three-body wave function \cite{gar03}.
The energy order of the $3/2^-$ and $5/2^-$ states is reversed and
the results are therefore questionable. The schematic model gives a larger
value for the direct decay, while sequential is forbidden by energy
conservation. The ratio between the outer and inner turning points is 
taken equal to 7 (taken from the effective potential in Fig.~\ref{fig7}),
and the knocking rate again equal to 3 MeV/$\hbar$. The value obtained
is larger than in the WKB calculation and the experiment because the strong
interaction substantially increases the barrier width due to the virtual
decay mechanism, where the two-body state in the decay is populated up
to very large distances.

\begin{figure}
\begin{center}
\vspace*{-1.1cm}
\epsfig{file=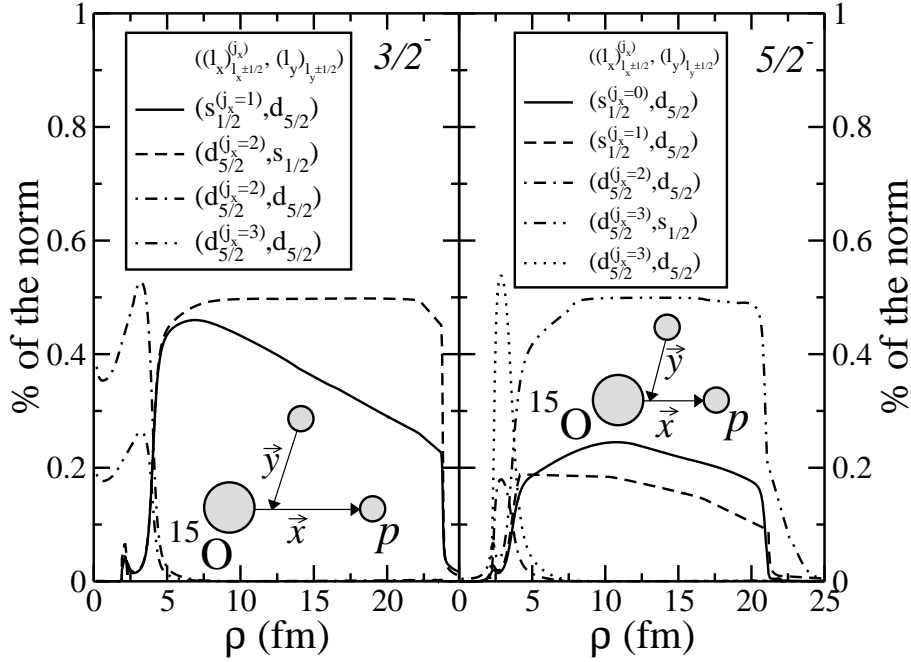,scale=0.5, angle=270}
\end{center}
\caption{The fraction of different components for the dominating adiabatic
potentials as function of $\rho$ for $^{17}$Ne($\frac{3}{2}^-$) (left)
and $^{17}$Ne($\frac{5}{2}^-$) (right).  The notation is as in 
Fig.~\ref{fig4}. }
\label{fig9}
\end{figure}

Our estimates indicate that the decay paths via the $s$ or
$d$-states of $^{16}$F are of comparable importance. The lowest
adiabatic potential shown in Fig.~\ref{fig8} contains the most
favorable combination obtained by the condition of adiabatic energy
minimization.  The contributions from the different components are
revealed by the probabilities shown in Fig.~\ref{fig9}.  Below the
barrier the $^{16}$F $d$-state remains steady on about 50\% while the
$s$-state population in the dominating adiabatic potential decreases
from 50\% down to about 20\% as $\rho$ increases up to 25~fm.  These
results demonstrate first the virtual sequential decay mechanism and
second that one adiabatic potential by clever coherent superposition
in itself simultaneously is able to describe two different sequential
decays via two different $^{16}$F states. The important channels are
already included in one adiabatic potential.

The $5/2^-$ resonance in $^{17}$Ne is energetically able to decay
sequentially via the two lowest states of $^{16}$F, see
Fig.~\ref{fig10}, but in both cases the centrifugal barrier must then
correspond to angular momentum 2.  We show the adiabatic potentials in
Fig.~\ref{fig7} where we find two low-lying potentials with long
relatively flat regions at intermediate distances. The energies
correspond almost to a constant two-body energy combined with the
decreasing Coulomb and centrifugal potentials arising from the emitted
proton.  The resonance wave function has probabilities of 77\% and
17\% on the first and second of these potentials, respectively.  The
probability distribution for the dominating component is shown in
Fig.~\ref{fig8}.  The one-peak distribution at small $\rho$ develops
into two ridges at small and large proton-core distances. This is
again the signature of a sequential decay.  The corresponding WKB
width is found to be $\simeq 1.3 \times 10^{-10}$~MeV (table~\ref{tab1})
again competing with $\gamma$-decay.  In comparison \cite{gri03} obtained 
$\simeq 1.2 \times 10^{-11}$~MeV and $\simeq 5 \times 10^{-9}$~MeV for direct 
and sequential decay, respectively.  These results are comparable to our
estimate of the width, i.e. 10 (2) times smaller (larger) for direct
(sequential) decay. 

The effective potentials in Fig.~\ref{fig7} suggest a ratio between
the outer and inner turning points of around 5 for the $\frac{5}{2}^-$
resonance. The values of the widths given in table~\ref{tab1} corresponding 
to the schematic calculations have been obtained using this value for the
ratio of turning points. The sequential width by proton emission has been 
computed assuming decay through the lowest resonance in $^{16}$F. The widths 
obtained 
in this way for the direct and sequential decay are very similar, and a not 
very significant change in the parameters (for instance the knocking rate
that is taken equal in both cases) could make one width or the other
to be the largest. So, this estimate does not permit to determine clearly
in this case which of the decays is preferred. Again the schematic model 
gives relatively large widths because the strong interaction in the realistic 
calculations maintain the two-body resonance populated in the intermediate 
step up to very large distances. This mechanism is the same for allowed and 
virtual sequential decay and in both cases the barrier width is significantly
increased.

The intermediate $^{16}$F two-body state could be either $0^-$ and
$1^-$ as allowed by energy conservation, or it could be $2^-$ and
$3^-$ if virtual sequential decay is favored as for the $3/2^-$ state,
see Fig.~\ref{fig10}.  Neither $0^-$ nor $1^-$ can be combined with an
$s$-wave of the emitted proton to produce the initial $5/2^-$
state. Then there seems to be no reason to choose the more expensive
path via the $1^-$ state.  On the other hand both the $2^-$ and $3^-$
states can combine with an $s$-wave and produce $5/2^-$. For these the
least expensive combination is then the $2^-$ state.  Thus we can
expect coherent contributions from sequential decay through the $0^-$
state and virtual sequential decay through the $2^-$ state.

The optimum combination is found by the dominating adiabatic potential
and the corresponding components are shown in Fig.~\ref{fig9} as
functions of $\rho$.  As for the $3/2^-$ state below the barrier the
$d$-state of $^{16}$F remains steady on about 50\% while the two
$s$-state populations each of about 20\% have a tendency to decrease
with $\rho$.  The sudden decrease to zero at about 22~fm is due to the
crossing of the two adiabatic potentials. The probabilities would
otherwise smoothly continue on the other potential.

The two $7/2^-$ and $9/2^-$ resonances in Fig.~\ref{fig10} are
probably not as clean three-body states since one core excitation with
an appropriate angular momentum and parity also is energetically
favorable.  Ignoring such a core-contribution three-body computations
show smaller and thinner barriers due to the higher energies.  The
flat regions under the barriers are no longer present.  On the other
hand the probability distributions behave qualitatively as for the two
lower-lying resonances. This again then strongly indicates sequential
decays. The larger angular momenta now require two $d$-states and the
intermediate $^{16}$F states must then be either $2^-$ or $3^-$. The
Coulomb and centrifugal barrier potentials are the same for both these
cases and therefore expected to provide roughly equal contributions.
Also for these resonances the short-range interaction is crucial for
the behavior of the dominating adiabatic potentials.  The reason is
again that one proton can be emitted while the attraction is exploited
in the remaining $^{16}$F state. In this way the short-range
interaction is decisive for the decay mechanism

\section{Summary and conclusions} 

As in the companion paper \cite{gar05} we investigate the decay of
low-lying continuum states into three-particle final states.  We focus
on the illuminating relation between the partial three-body decay
widths and the decay mechanisms.  We assume that the three fragments
are formed before entering the barrier at sufficiently small distances
to allow the three-body treatment.  This is analogous to the two-body
$\alpha$-decay and seems to be a reasonable working hypothesis which
however for some structures may turn out to be inaccurate.  We are
then left with a three-body problem consisting of the three
interacting particles eventually appearing in the final state. By
definition the energy is positive measured relative to the break-up
threshold.  The necessary technique must therefore be able to describe
three-body continuum structure and in particular three-body
resonances.  It is worth emphasizing that the initial state can be a
complicated many-body state with three-body decay as one, more or
less, probable decay channel.

The hyperspherical adiabatic expansion method is an efficient tool to
compute bound states and resonances when combined with the complex
rotation method.  This means that rather accurate calculations of
three-body resonance widths are available. Furthermore, one adiabatic
potential is often dominating a given resonance and therefore in
itself providing a reasonable estimate of the width.  The wave
function along this adiabatic potential then presents a continuous
connection between initial and final states. It can be viewed as a
classical path providing the largest contribution to the width. In
principle other potentials should also be included to go beyond the
classical treatment of this coordinate. This is achieved by the
complex rotation method.  Even with one adiabatic potential all other
coordinates are treated quantum mechanically correct by including a
distribution of amplitudes coherently adding up to produce this
individual adiabatic potential.

We introduce the short-range interaction in realistic calculations.
Direct decay is demonstrated for $^{6}$He.  The intermediate
configurations turn out to be at higher energies than along the direct
path.  The Coulomb dominated decays, exemplified by two 0$^+$ states
at different energies in $^{12}$C, are predicted to be either
sequential (low energy) or direct (high energy).  Two-proton decays of
the proton dripline nucleus $^{17}$Ne is strongly sequential for the
two lowest excited $3/2^-$ and $5/2^-$ resonances. However, the
intermediate two-body state in $^{16}$F has too high an energy to be
populated from decay of the $3/2$-state. Still the mechanism is
clearly sequential by keeping proton and core spatially close up to very
large average distances. This can only be achieved by exploiting the
attractive two-body short-range interaction. The process is called
virtual sequential decay.  It is then not surprising that also the
$5/2$-state favors sequential decay.

In conclusion we have demonstrated how to classify and characterize
various decay mechanisms, i.e. sequential, direct and virtual
sequential.  We illustrated by elaborate realistic computations. The
intermediate configurations strongly indicate which basis is most
efficient in analyses and interpretations of experimental data.  The
three-body structure can be fully explained in a complete basis chosen
as the intermediate two-body sequential decay states multiplied by
distorted waves for the third particle. This does not imply that all
decays are sequential. Another complete three-body basis like the
adiabatic states can also be used and in this paper demonstrated to
include both aspects in a natural way.

\end{document}